\documentclass[iop, apjl, twocolappendix, twocolumn]{aastex631}
\usepackage{graphicx}
\usepackage{bm}
\usepackage{amsmath,amssymb}
\usepackage{natbib}
\usepackage{booktabs}
\usepackage{color}
\usepackage{units}
\usepackage[normalem]{ulem}

\usepackage[normalem]{ulem}
\newcommand{\mtov}{\ensuremath{M_{\rm TOV}}}
\newcommand{\program}[1]{\textsc{#1}}
\newcommand{\citeg}[1]{\citep[e.g.,][]{#1}}
\newcommand{\grb}{GRB 201006A}
\newcommand{\be}{\begin{equation}}
\newcommand{\ee}{\end{equation}}

\begin{document}
\title{The origin of the coherent radio flash potentially associated with GRB 201006A}

\author[0000-0003-2700-1030]{Nikhil Sarin}
\affil{Oskar Klein Centre for Cosmoparticle Physics, Department of Physics,
Stockholm University, AlbaNova, Stockholm SE-106 91, Sweden}
\affil{Nordita, Stockholm University and KTH Royal Institute of Technology \\
Hannes Alfvéns väg 12, SE-106 91 Stockholm, Sweden}
\author[0000-0002-6714-5429]{Teagan A. Clarke}
\affil{School of Physics and Astronomy, Monash University, VIC 3800, Australia}
\affil{OzGrav: The ARC Centre of Excellence for Gravitational-Wave Discovery, Clayton, VIC 3800, Australia}

\author[0009-0000-7037-1809]{Spencer J. Magnall}
\affil{School of Physics and Astronomy, Monash University, VIC 3800, Australia}
\affil{OzGrav: The ARC Centre of Excellence for Gravitational-Wave Discovery, Clayton, VIC 3800, Australia}

\author[0000-0003-3763-1386]{Paul~D. Lasky}
\affil{School of Physics and Astronomy, Monash University, VIC 3800, Australia}
\affil{OzGrav: The ARC Centre of Excellence for Gravitational-Wave Discovery, Clayton, VIC 3800, Australia}

\author[0000-0002-4670-7509]{Brian D.~Metzger}
\affil{Theoretical High Energy Astrophysics (THEA) Group, Columbia University, New York, NY 10027, USA}
\affil{Department of Physics and Columbia Astrophysics Laboratory, Columbia University, New York, NY 10027, USA}
\affil{Center for Computational Astrophysics, Flatiron Institute, 162 5th Ave, New York, NY 10010, USA} 

\author[0000-0002-9392-9681]{Edo Berger}
\affil{Center for Astrophysics | Harvard \& Smithsonian, 60 Garden St. Cambridge, MA 02138, USA}

\author[0000-0002-5519-9550]{Navin Sridhar}
\affil{Theoretical High Energy Astrophysics (THEA) Group, Columbia University, New York, NY 10027, USA}
\affil{Department of Astronomy, Columbia University, New York, NY 10027, USA}
\affil{Cahill Center for Astronomy and Astrophysics, California Institute of Technology, Pasadena, CA 91106, USA}

\begin{abstract}
\citet{rowlinson23} recently claimed the detection of a coherent radio flash 76.6 minutes after a short gamma-ray burst. They proposed that the radio emission may be associated with a long-lived neutron star engine. We show through theoretical and observational arguments that the coherent radio emission, if real and indeed associated with GRB 201006A and at the estimated redshift, is unlikely to be due to the collapse of the neutron star, ruling out a blitzar-like mechanism. Instead, we show if a long-lived engine was created, it must have been stable with the radio emission likely linked to the intrinsic magnetar activity. However, we find that the optical upper limits require fine-tuning to be consistent with a magnetar-driven kilonova: we show that neutron-star engines that do satisfy the optical constraints would have produced a bright kilonova afterglow that should already be observable by the VLA or MeerKAT (for ambient densities typical for short GRBs). Given the optical limits and the current lack of a kilonova afterglow, we instead posit that no neutron star survived the merger, and the coherent radio emission was produced far from a black hole central engine via mechanisms such as synchrotron maser or magnetic reconnection in the jet---a scenario consistent with all observations. We encourage future radio follow-up to probe the engine of this exciting event and continued prompt radio follow-up of short GRBs.
\end{abstract}

\section{\label{sec:intro}Introduction}
The spectacular multi-messenger observations of a binary neutron star merger coincident with a short gamma-ray burst (sGRB) GW170817~\citep{170817_discovery, Abbott_2017_multi} confirmed the long-argued association of binary neutron star mergers with short GRBs \citep{npp92,ber14}. Despite this watershed event and an ever-growing population of sGRB observations (e.g., \citealt{fbm+15}), key questions remain unanswered about the nature of the engine of sGRBs, and the potential precursor and postcursor emissions. 

The myriad of electromagnetic phenomena post-merger associated with a binary neutron star merger are intrinsically connected to the fate of the merger. The fate of the merger \citep[see][for a review]{Sarin2021_review} primarily depends on two quantities: the total mass of the binary and the Tolman-Oppenheimer-Volkoff (TOV) mass, \mtov{}, a property of the elusive equation of state. There is also a weak dependence on the spins of the progenitors and angular momentum profile~\citeg{Rosswog2023}. A merger with a remnant mass below $\approx 1.2\times \mtov{}$ is expected to form a long-lived ($\geq \unit[10]{s}$) neutron star engine as opposed to a more prompt formation of a black hole. 

Long-lived neutron star engines have been argued to result in a plethora of electromagnetic emissions potentially associated with binary neutron star mergers and sGRBs, such as the internal and external plateaus seen in X-ray afterglows~\citeg{Troja07, rowlinson10, Rowlinson_2013}, engine-driven kilonovae~\citeg{Metzger2014b,Metzger2014c}, and kilonova afterglows that are brighter than ones without an engine~\citeg{Metzger2014_grb}. They may also be progenitors of fast radio bursts~\citep[FRBs; e.g.,][]{Lyubarsky_2014, Metzger_2017, Lu_2018, Sridhar+21}. 
These associations and the potential to probe nuclear matter and neutron-star astrophysics in the early lives of neutron stars make these objects exciting laboratories.

\begin{figure*}[t]
    \centering
    \includegraphics[width=1.0\textwidth]{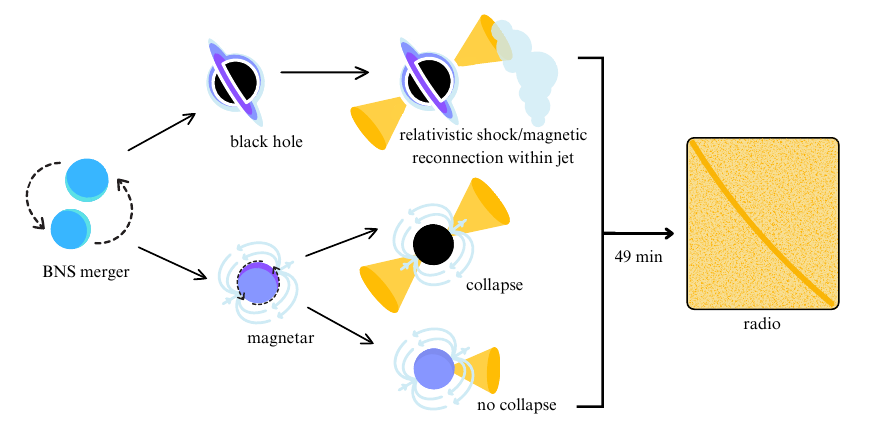}
    \caption{Schematic showing the possible outcomes of a binary neutron star merger that could create the coherent radio emission seen in GRB 201006A after 49\,min (in the source frame at $z=0.58$).}
    \label{fig:schematic}
\end{figure*}

GRB 201006A~\citep[][]{gropp20} was a sGRB with a T$90 \leq \unit[0.5]{s}$, detected by the \textit{Neil Gehrels Swift Observatory}, with an X-ray afterglow, and a claimed associated radio flash $76.6$ minutes (in the observer's frame) after the prompt emission, despite a large offset of $27\pm 7''$ from the X-ray position \citep{rowlinson23}. 
No host-galaxy has been identified~\citep{Fong_2022}, but a dispersion-measure analysis of the radio flash placed this source at a redshift of $z=0.58 \pm 0.06$, comparable to other sGRBs. We note that the redshift measurement is not certain and relies on the tentative dispersion measurement and assumptions of the dispersion measure contributions of the host galaxy.
The X-ray afterglow of GRB 201006A has been interpreted to show signatures of energy injection from a nascent neutron star, with an external plateau reminiscent of other sGRBs that are claimed to have a long-lived neutron-star engine. 

The nature of the coherent radio emission (if it is real) is unclear, with \citet{rowlinson23} suggesting the emission is either linked with the delayed collapse of the neutron star into a black hole or the intrinsic activity of a stable neutron star. In the event of a delayed collapse, the coherent radio emission was suggested to be produced via the blitzar mechanism~\citep{falcke14}, while in the latter scenario, the radio emission was postulated to be generated via magnetar activity akin to the recent FRB-like emission from the Galactic magnetar, SGR\,1935+2154~\citep{Chime_2020, Margalit_2020}.

Coherent radio emission analogous to FRBs has previously been claimed in association with the binary neutron star merger GW190425~\citep{Abbott2020_1902425}, where again, a blitzar mechanism was invoked for FRB 190425~\citep{CHIME/FRBCollaboration2021}. This event occurred $\unit[2.5]{hrs}$ post-merger, and was independently identified by CHIME to be within the sky localization of GW190425~\citep{Moroianu2023}. 
Although this event's coincidence was ultimately deemed not to be statistically significant~\citep{Abbott2023_lvkchime,Bhardwaj2023, MaganaHernandez2024}, it raised many theoretical considerations into such associations for the future. For example, (1) the radio bursts cannot escape the ejecta unless viewed through the polar funnel until the ejecta becomes optically thin, which can take up to two weeks post-merger~\citep{margalit18_frbengines, margalit2019_frbbns, Bhardwaj2023}, and (2) non-detections at optical wavelengths that rule out magnetar-driven kilonovae~\citep{Smartt2024}.

In this \emph{Letter}, we explore the possible progenitor of the radio flash from GRB 201006A and the nature of its engine. In Sec.~\ref{sec:association}, we explore how confidently the radio flash can be associated with \grb{}. We also focus on theoretical considerations, in particular in  Sec.~\ref{sec:collapse}, we examine the claim of a neutron-star engine that collapsed $\approx \unit[77]{min}$ (in observer frame) after the sGRB to produce the coherent radio flash. In Sec.~\ref{sec:stableNS}, we review the scenario where the engine is an indefinitely stable neutron star, including predictions of optical emission from a magnetar-driven kilonova and radio afterglow that are in tension with constraints. 
We also explore in Sec.~\ref{sec:just-GRB} the scenario that a black hole engine survived this merger, with a synchrotron-maser or magnetic reconnection as the origin for the radio flash, a scenario consistent with all observations. 
We conclude in Sec.~\ref{sec:discussion} by reviewing all possibilities (which we summarise in Fig.~\ref{fig:schematic}), the implications of this event, and highlight future observations to ascertain the engine and source of the coherent radio emission. 

\section{How confident is the association?}\label{sec:association}

The veracity of the coherent radio source, its dispersion measure (and therefore redshift measurement), and its association with GRB\,201006A are not robust~\citep{rowlinson23}. 
The radio flash detected in LOFAR is $27\pm 7$ arcsec offset from the GRB location, and the positions are thus inconsistent at the $4\sigma$ level. \citet{rowlinson23} argued that the probability of an unrelated transient occurring within 40 arcsec of the sGRB is $\leq 10^{-6}$. This raises the question of whether the radio flash and gamma-ray burst are actually coincident.  

The \textit{a priori} odds of whether an FRB-like event detected in LOFAR can be considered coincident with a sGRB detected by \textit{Swift} is~\citep{Ashton2018}
\begin{equation}
\pi_{C/R} = \frac{R_{\rm{FRB, sGRB}}}{R_{\rm FRB}R_{\rm sGRB} T_{\rm obs}}, 
\end{equation}
where $R_{\rm FRB, sGRB}$ is the rate of FRBs from sGRBs, $R_{\rm FRB}$ is the rate of FRB-like events detected by LOFAR, $R_{\rm sGRB}$ is the rate of short gamma-ray bursts detected by \textit{Swift}, and $T_{\rm obs}$ is the co-observing time i.e., the window of time searched for every event. 

The rate of FRB-like transients per year out to a redshift of $0.58$ in LOFAR is $\unit[2.2 \times 10^{6}]{yr^{-1}}$, Following the LOFAR upper limit on FRB-like transients of $\unit[134]{Gpc^{-3}day^{-1}}$\citep{terVeen2019}, while the rate of sGRBs observable by \textit{Swift} is $\unit[10]{yr^{-1}}$.
Conservatively (for the purposes of this calculation) assuming that all short gamma-ray bursts progenitors can produce FRB-like emission and a $T_{\rm obs} = \unit[2]{hrs}$ co-observing time~\citep{rowlinson23} gives a prior odds of $2\times 10^{-3}$, i.e., \textit{a priori} one in approximately $500$ coincident events would be from the same astrophysical source, with a chance coincidence approximately $500$ times more likely than a real coincidence. 

This above calculation is based only on \textit{a priori} knowledge and excludes the other observed properties of \grb{} and the potentially associated radio flash, which could raise or further decrease the confidence of this association. For example, a potential overlap in the properties of the radio flash and \grb{} such as the distance and progenitor properties, could raise or decrease the coincidence significance of this event~\citep[see][for details]{Sarin2022_nemo, Sarin2023_missed}. 
However, without detailed models of estimating such properties from the sGRB or radio emission, we leave such considerations to future work. Instead, we now turn towards theoretical considerations for the nature of the radio flash's origin, assuming the radio flash and sGRB are indeed correctly associated.
\section{A collapsing neutron star remnant?}\label{sec:collapse}
\citet{rowlinson23} posit that the binary neutron star merger responsible for GRB 201006A could have created a neutron star remnant supported against gravitational collapse through rigid body rotation. Such a remnant must have collapsed in $\approx \unit[49]{min}$ post-merger (in the rest frame; assuming $z=0.58$) to form a black hole, with the coherent radio emission produced according to the Blitzar model~\citep{falcke14}. 
In this Section, we show that this model is disfavoured for two reasons: first, the theoretically and observationally derived collapse times of millisecond magnetars should be shorter than this 49-minute timescale, and second, the X-ray data are not consistent with this model. 

\begin{figure*}[t]
    \centering
\includegraphics[width=1.0\textwidth]{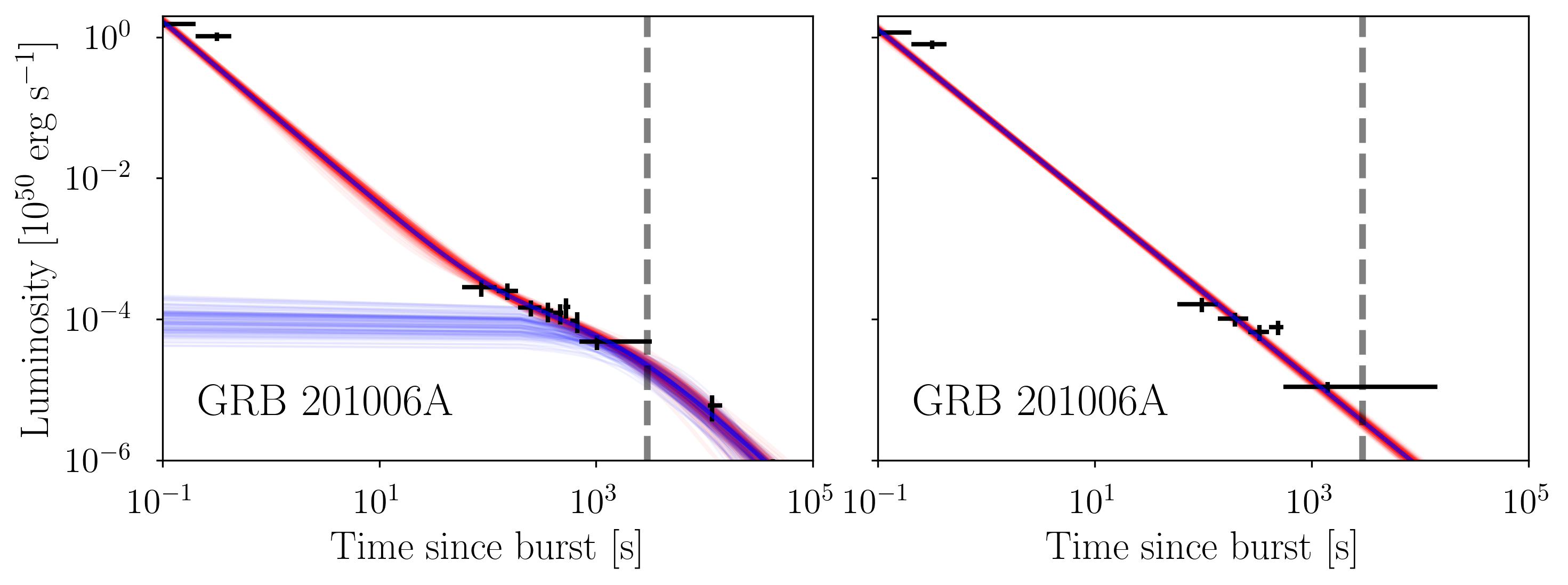}
    \caption{X-ray afterglow data of GRB 201006A (assuming $z=0.58$) with binning consistent with \citet{rowlinson23} in the left panel and \textit{Swift} automatic binning on the right panel. The left panel shows 100 random draws (red curves) from the posterior of a fit with a magnetar and power-law model, while the right panel shows the same but with just the power-law model. In the left panel we also include the inferred luminosity of the magnetar itself (blue curves). The gray vertical line indicates the time of the radio flash.}
    \label{fig:xraylcs}
\end{figure*}

\subsection{Theory: 49 minutes is too long}
Models of neutron star spin-down show supramassive neutron stars born in the aftermath of binary neutron star mergers collapse within $\approx \unit[4\times10^4]{s}$~\citep{ravi14}. These models are overly simplisitic, assuming spin down is entirely due to dipole magnetic-field radiation, while also ignoring gravitational-wave emission and assuming purely hadronic equations of state. 
Dipolar spin down is inconsistent with observed braking indices for most neutron stars in our Galaxy~\citep{Parthasarathy2020}, and unlikely to be true for newly born neutron stars, which are expected to have significant gravitational-wave emission~\citeg{anderson21} and complex magnetic-field morphologies~\citep{melatos97}.
Most physically-motivated modifications to simple dipole spin-down models are consistent with predicting shorter collapse times, which includes additional angular-momentum loss through gravitational-wave emission~\citep[e.g.,][]{gao16} or different braking indices expected from including more detailed physics, such as more complex magnetic-field structures (e.g., due to the effects of fall-back accretion expanding the open field lines; \citealt{Metzger+18}) or evolution of the magnetic-field axes~\citep[e.g.,][]{melatos97, lasky17}. 
As with detailed models of spin-down, consideration of more general classes of the equation of state also serves to drive the distribution of collapse times to shorter timescales, by up to an order of magnitude from the $\approx \unit[4\times10^4]{s}$ timescale above ~\citep{li16}. The combination of these effects makes it difficult to reconcile the long collapse timescale required by the Blitzar-model origin for the radio flash in GRB 201006A. 

A critical look at the astrophysical population from X-ray afterglows of sGRBs also suggests a collapse time distribution inconsistent with $\approx\unit[49]{min}$~\citep[e.g.,][]{fan13,ravi14,Sarin_2020}. In particular, the $18$ systems all interpreted to have neutron star remnants that collapse into a black hole, do so on timescales shorter than $\unit[10^3]{s}$, with the majority collapsing within $\unit[300]{s}$~\citep{Sarin_2020}. This is in significant tension with the $\unit[49]{min}$ timescale required for a Bliztar model origin of the radio flash in GRB 201006A. We note that while the dispersion measure and, therefore, redshift measurement of GRB 201006A is tentative, no probable sGRB redshift would make the observed $\unit[77]{min}$ timescale consistent with the above considerations.

\begin{figure*}[t]
    \centering
    \includegraphics[width=1.0\textwidth]{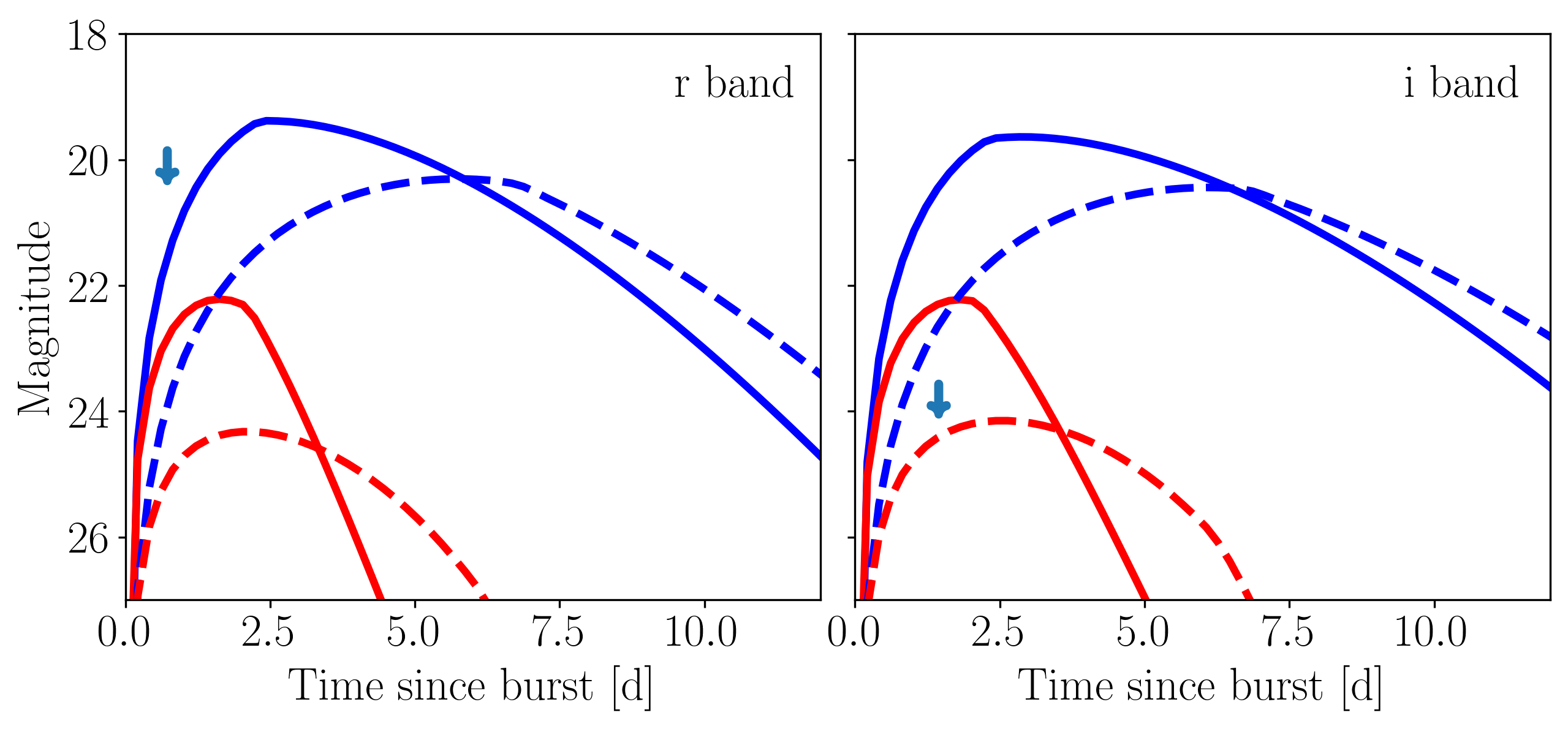}
    \caption{R-band (left) and i-band (right) lightcurves for a magnetar-driven kilonova for two different types of magnetar engines at $z=0.58$: one where the bulk of the rotational energy thermalizes the ejecta (blue curves) and another where the rotational energy is instead used in accelerating the ejecta (red curves). The arrows indicate the upper limit set by follow-up observations of GRB 201006A. The solid curves are for an engine with a spin period of $\unit[0.7]{ms}$ whereas the dashed curves are for an engine with a slower initial spin period of $\unit[2]{ms}$.}
    \label{fig:opticallcs}
\end{figure*}

\subsection{Observations: no internal plateau in X-ray afterglow }
X-ray afterglows of several sGRBs have a plateau-like feature followed by a sharp drop in luminosity~\citep{Rowlinson_2013}. These ``internal plateaus" are often interpreted as the signature of a neutron star engine that collapses into a black hole (the collapse synchronous with the sharp drop in luminosity). 

To investigate whether the collapse time is consistent with the radio flash, we fit the X-ray afterglow data of GRB 201006A from \citet{rowlinson23} with a collapsing neutron star model~\citep{Sarin_2020} using \program{Redback}~\citep{sarin23_redback}, with the \program{pymultinest} nested sampler~\citep{Feroz2009, Buchner2016} wrapped with \program{Bilby}~\citep{Ashton2019a, RomeroShaw2020a}. 
We do not find any scenario where the neutron star engine contributes to the plateau emission and collapses (with a correspondingly sharp drop in luminosity) at $\unit[49]{min}$ while explaining the entire X-ray light curve.

\citet{rowlinson23} suggest that such a signature of collapse (i.e., a plateau with a sharp drop in luminosity) could be hidden by the synchrotron emission from the jet interacting with the interstellar medium. 
However, explaining the shallower decay at $\approx \unit[10^{3}]{s}$ suggested by their data from a jet would require complex jet physics, such as the evolution with time of microphysical parameters or imposing that the jet was viewed off-axis~\citep{Sarin2019, Beniamini2020}, in tension with the strong gamma-ray emission seen in GRB 201006A. 

\section{An indefinitely stable neutron star}\label{sec:stableNS}
\subsection{X-ray constraints}
A radio flash does not necessarily require a neutron-star remnant to have collapsed. Newly born magnetars are one of the front-running progenitors of FRBs and many mechanisms involving newly born magnetars, do not require collapse~\citep[e.g.,][]{katz_2016, Beloborodov_2017}.  

In this Section, we consider the possibility that the potential neutron-star merger responsible for GRB 201006A produced an indefinitely stable neutron-star remnant. This is possible provided that the remnant mass is below $\mtov{}$.

In the left-hand panel of Fig~\ref{fig:xraylcs} we show the X-ray data of GRB 201006A in black as observed by the \textit{Swift} observatory, and binned similarly as in~\citet{rowlinson23} (we discuss the effects of binning, as shown in the right panel, in Section \ref{sec:just-GRB}). 
The X-ray flux data are converted into luminosity assuming a redshift of $z=0.58$, inferred from the tentative dispersion of the coherent radio source. 
The red curves show $100$ random draws from the posterior for a fit with the magnetar model with variable braking index~\citep{lasky17}. This model assumes that some constant fraction of the bare (shown as blue curves) spin-down luminosity of the magnetar is converted into X-rays alongside a power-law to model the emission from jet. These red curves are in agreement with the observations. 
We infer the spin-down damping timescale 
$\tau = {1477}_{-627}^{+1721}$~s ($68\%$ credible interval), broadly consistent with the timescale for the radio flash. However, we note that the spin-down timescale may not be a relevant quantity for the origin of the coherent radio emission if the emission is produced through magnetic activity instead of through spin-down, which may also be insufficient to explain the energetics of fast radio bursts. 
The braking index is not well measured (relative to the prior) with an uninformative posterior of $n = {5.00}_{-1.68}^{+1.34}$ ($68\%$ credible interval), i.e., consistent with dominant spin down from either a vacuum dipole or gravitational-wave emission. 

\subsection{Optical constraints}
The X-ray data of GRB 201006A is in agreement with the prediction of the magnetar model. However, if GRB 201006A indeed harboured an indefinitely stable neutron-star engine, then other signatures of the neutron star must also satisfy multi-wavelength constraints. The cleanest probe to infer the engine for GRB 201006A would have been a direct detection of gravitational waves from the nascent neutron star. While gravitational-wave detectors were operating at this time, detection at a redshift of $z=0.58$ would not have been possible, even for third-generation observatories~\citep{Dallosso_2015, lasky16, sarin2018}.

The location of GRB 201006A was observed by optical telescopes in the immediate aftermath of the sGRB. In particular at $\unit[0.7]{d}$ in r-band~\citep{GCN-rband_GROWTH-India} and at $\unit[1.44]{d}$ in i-band~\citep{GCN-Lowell_iband}. If the remnant of GRB 201006A is an indefinitely stable neutron star, then these optical upper limits should also be consistent with a magnetar-driven kilonova. 
To investigate this, we calculate magnetar-driven kilonova light curves at $z=0.58$ for both filters following the magnetar-driven kilonova model in~\citet{Sarin_2022} implemented in \program{Redback}. 
These light curves are shown in Fig.~\ref{fig:opticallcs} (r-band in the left panel and i-band in the right panel). 
Given the large uncertainty in the parameters of the model, we show two characteristic magnetar engines with ejecta parameters comparable to AT2017gfo~\citep[e.g.,][]{Villar2017}. 
We note that we include gravitational-wave emission from a magnetic-field deformation for the two characteristic engines. In particular, the total ejecta mass is $\unit[0.07]{M_{\odot}}$, with a bulk velocity of $\unit[0.2]{c}$ and gray opacity $\unit[1]{g^{-1}~cm^2}$. For the neutron-star engine, we assume two initial spin periods of $\unit[0.7]{ms}$ (solid curves) and $\unit[2]{ms}$ (dashed curves), and a $\unit[11]{km}$ radius. 
The blue curves show an engine whose magnetic field configuration directs a large fraction of the neutron star spin-down energy toward thermalizing the ejecta. On the other hand, the red curves are for a magnetic field configuration in which the bulk of the system's spin-down energy budget is lost in accelerating the ejecta. 

As Fig.~\ref{fig:opticallcs} indicates, the i-band upper limit provides much of the constraining power. The i-band upper limit indicates that the data is better explained by a slowly rotating neutron star engine that utilizes the bulk of its rotational energy budget to accelerate the ejecta and does not thermalize the ejecta. However, such an acceleration is also not without consequence, as the acceleration will present itself in the kilonova afterglow. 

In Fig.~\ref{fig:radiolcs}, we show the radio $\unit[1.4]{GHz}$ kilonova afterglow for a neutron-star engine with parameters corresponding to the same blue and red model light curves as above, following the kilonova afterglow model in~\citet{Sarin_2022} implemented in \program{Redback}. The thickness of the band indicates the uncertainty in the kilonova afterglow from two orders of magnitude uncertainty in the ambient interstellar medium density ($10^{-4}-10^{-2}~\unit{cm^{-3}}$)~\citeg{fbm+15}. 
The horizontal solid grey band marks the detection limit of the Very Large Array~\citep[VLA;][]{VLA2020}, while the dashed grey band corresponds to the detection limit of MeerKAT~\citep{MeerKAT}. 
Unlike the kilonova, which requires the ejecta to be thermalized, the kilonova afterglow is sensitive to the bulk kinetic energy. 
As a result, engines where the bulk of the rotational energy is lost in accelerating the ejecta (red curves), produce a much brighter kilonova afterglow, one that is likely detectable now in the VLA for interstellar medium densities comparable to other sGRBs. The MeerKAT telescope sensitivity may be sufficient to probe both types of engines at the nominal redshift of $z=0.58$. We return to this point in Sec.~\ref{sec:discussion}.

\section{No magnetar at all} \label{sec:just-GRB}
\subsection{Observational constraints}
A scenario not explored in detail by \citet{rowlinson23}, but worth investigating further, is: what if no magnetar was born in the merger? We argue that the data is best explained by this scenario and requires the least amount of fine-tuning.  

We first consider the X-ray afterglow. As alluded to in Sec.~\ref{sec:stableNS}, to interpret the X-ray afterglow as evidence for energy injection, we follow the binning prescription for the lightcurve used in \citet{rowlinson23}. 
However, this choice of binning is not the default binning strategy recommended by the Swift Data Centre. In particular, to create the binned lightcurve shown in the left-panel of Fig.~\ref{fig:xraylcs}, we had to reduce the minimum significance allowed for each bin to $1\sigma$ (cf., default of $3\sigma$) and reduce the minimum counts per bin to $15$ (cf., default of $30$). 
In the right panel of Fig.~\ref{fig:xraylcs}, we show the binned lightcurve following the automatic binning strategy. The automatically binned lightcurve has a markedly different shape. 
In the same panel, we show a fit using only a power-law model i.e., simplified modelling of the synchrotron emission produced by the jet interacting with the interstellar medium~\citep{sari99}. We also tried a model with a magnetar. However, the data overwhelmingly favours the simple power-law fit instead of showing hints of energy injection from a stable neutron star. This highlights that the interpretation of the X-ray afterglow of GRB 201006A is highly dependent on the choice of binning, and reinforces the need to perform these analyses without binning~\citeg{sarin2021_xt1}.

\begin{figure}
    \centering  \includegraphics[width=0.95\columnwidth]{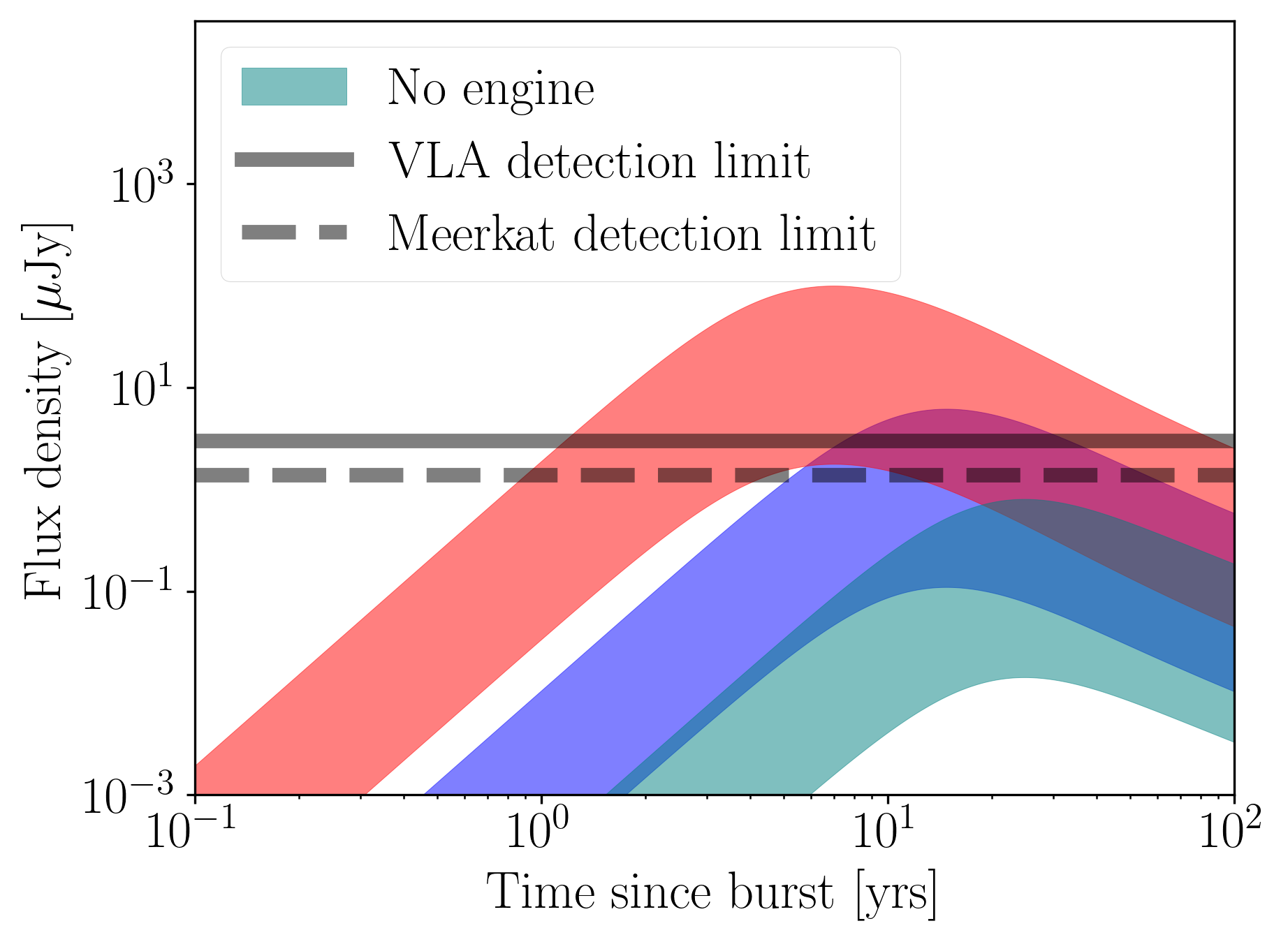}
    \caption{Kilonova afterglow at $\unit[1.4]{GHz}$ for the same magnetar parameters as in Fig.~\ref{fig:opticallcs}, the width of the bands indicate the uncertainty in the kilonova afterglow for ambient interstellar medium densities between ($10^{-4}-10^{-2}~\unit{cm^{-3}}$). The horizontal solid and dashed bands indicate the detection limits of the VLA and Meerkat, respectively.}
    \label{fig:radiolcs}
\end{figure}

Optical constraints also disfavour a magnetar-driven kilonova. 
Furthermore, these neutron star central engines should produce a bright radio remnant that would be detectable now in the VLA or with MeerKAT. To our knowledge, no such radio remnant has yet been detected in the VLA all-sky survey. Based on the above results, we interpret the GRB 201006A to be powered by a simple, canonical gamma-ray burst without energy injection. Here, the X-ray afterglow can be explained as a power-law following expectations of synchrotron emission from a jet interacting with the ambient interstellar medium~\citep{sari99}. Meanwhile, the optical and radio constraints are easily satisfied given the lack of an engine makes both significantly dimmer (see teal curves in Fig~\ref{fig:radiolcs}).

\subsection{Physical constraints}

The optical depth to an FRB-like burst due to the free ions and electrons from the near-isotropic disk wind ejecta following a neutron star merger is \citep{Rybicki&Lightman_79},
\begin{equation}\label{eq:varepsilon_min}
\tau_{\rm ff} = \alpha_{\rm ff}R_{\rm ej}(t) \simeq 0.018Z^2\nu^{-2}T_{\rm ej}^{-\frac{3}{2}}n_{e}n_{\rm ion}\bar{g}_{\rm ff}R_{\rm ej}(t).
\end{equation}
Here, $Z$ is the atomic number, $\nu$ is the frequency, $T_{\rm ej}$ is the ejecta temperature, $n_{e}$ and $n_{\rm ion}$ are the number density of electrons and ions respectively, $\bar{g}_{\rm ff}\approx1$ is the Gaunt factor, and $R_{\rm ej}$ is the ejecta radius. 
We assume the ejecta to be populated by the typical light r-process elements (e.g., Xenon, with $Z=54$ and atomic mass number $A=131$), with a speed of $v_{\rm ej}=0.1\,c$, a temperature of $T_{\rm ej}=10^5$\,K (at $t=49$\,min), and the mass in the ejecta to be $M_{\rm ej}=10^{-3}\,M_{\odot}$ \citep{Metzger_19}. The radial extent of the ejecta at $t=49$\,min post-merger is $R_{\rm ej}\sim v_{\rm ej}t\approx10^{13}$\,cm.  The ion number density is $n_{\rm ion}=3M_{\rm ej}/(4\pi R_{\rm ej}^3Am_{\rm p})$, with a corresponding electron number density of $n_{\rm e}\sim n_{\rm ion}Z$. At $\nu=\unit[1]{GHz}$, the optical depth is, 
\begin{align}
\tau_{\rm ff} =
& \approx 10^{16} \left(\frac{\nu}{{\rm GHz}}\right)^{-2} \left(\frac{T_{\rm ej}}{10^5\,{\rm K}}\right)^{-\frac{3}{2}} \left(\frac{v_{\rm ej}}{0.1\,c}\right)^{-5} \nonumber \\
& \times \left(\frac{M_{\rm ej}}{10^{-3}\,M_\odot}\right)^{2} \left(\frac{t}{49\,{\rm min}}\right)^{-5},
\end{align}
The free-free optical depth is too large for FRB-like emission to escape (until $\sim 54$ days post-merger for the above assumed parameters), even in a scenario where the GRB jet clears out a majority of the mass along the polar funnel~\citep{Zhang_14}. We note that this timescale could be shorter due to recombination into neutral gas, but not as short as the $\unit[49]{min}$ timescale required for GRB 201006A. Instead, we require alternative scenarios where the FRB is emitted at large distances away from the central engine.

The coherent radio emission could be produced at large distances $\gtrsim10^{13}$\,cm away from a newly formed black hole via a synchrotron maser~\citep{Metzger+19} or magnetic reconnection \citep{Lyubarsky_20}. Such a scenario would require the launch of a flare by the accreting black hole central engine, near the time of the coherent radio emission, which propagates out and creates a shock against the upstream jet material, producing radio emission through synchrotron maser instability. Alternatively, the flare can trigger magnetic reconnection within the striped magnetic fields in the jet and give rise to plasmoids that merge and produce a coherent radio flash. This scenario is similar to the proposed FRB generation mechanism from luminous accreting X-ray binaries~\citep{sridhar21}. 

\cite{rowlinson23} considered the possibility of a black hole remnant for GRB 201006A in which the FRB would be produced by synchro-Compton emission by the high-energy electrons accelerated due to the interaction between the magentized disk wind and the ambient medium~\citep{Usov&Katz_00}. They assumed that such an FRB emission (with a luminosity of $\sim10^{42}\,{\rm erg\,s^{-1}}$) should be associated with an X-ray flare with a luminosity of $\sim10^{47}\,{\rm erg\,s^{-1}}$ (for a redshift of $z=0.58$) from the highly elliptical parcel accreting onto the black hole, also presumably responsible for the FRB emission. \cite{rowlinson23} rejected the possibility of a black hole engine due to the lack of such an X-ray flare. 

However, in the scenario proposed by us \citep[\`{a} la][]{sridhar21}, the accretion rate onto the black hole required to produce an FRB with luminosity of $10^{42}\,{\rm erg\,s^{-1}}$ is $\dot{M}_\bullet\sim10^5\,\dot{M}_{\rm Edd}$ (see Fig.~2 of \citealt{Sridhar&Metzger_22}), where $\dot{M}_{\rm Edd}\equiv 0.1L_{\rm Edd}/c^2$ is the Eddington accretion rate corresponding to an accretion luminosity of $L_{\rm Edd}\sim10^{39}\,{\rm erg\,s^{-1}(M_\bullet/10\,M_\odot)}$, assuming a black hole of mass $M_\bullet=10\,M_\odot$. For 
$\dot{M}_\bullet/\dot{M}_{\rm Edd}\gg1$, the bolometric luminosity of the disk blackbody emission is $L_{\rm bol}=L_{\rm Edd}\left[1+\ln(\dot{M}_\bullet/\dot{M}_{\rm Edd})\right]$ \citep{Begelman+06}. Motivated by radiation magnetohydrodynamical simulations of super-Eddington accretion \citep{Sadowski&Narayan15}, we adopt an X-ray beaming fraction from a super-Eddington accreting funnel to be $f_{\rm b,X}\approx 10^{-3}(\dot{m}/10^{3})^{-1}$, and estimate the expected isotropic-equivalent X-ray luminosity to be, 
\begin{align} \label{eq:L_X}
L_{\rm X} &= \frac{L_{\rm bol}}{f_{\rm b,X}} \nonumber 
\\
&\approx 10^{45}\,{\rm erg\,s}^{-1}\left[1 + \ln \left(\frac{\dot{M}_\bullet}{10^5\,\dot{M}_{\rm Edd}}\right)\right]\left(\frac{\dot{M}_\bullet}{10^5\,\dot{M}_{\rm Edd}}\right).
\end{align}
This X-ray luminosity is consistent with the limit on X-ray flux set by \textit{Swift} (for $z=0.58$) at the time of the FRB emission. Furthermore, these hypotheses for the origin of the potentially associated coherent radio emission also alleviate the concern of how coherent radio emission is generated and propagates through the optically thick merger ejecta, as the emission is produced at large radii with significantly lower optical depth~\citep{margalit18_frbengines, margalit2019_frbbns, Bhardwaj2023}. We refer readers to \cite{sridhar21} for a detailed discussion on the expected FRB energetics, duration, and the contribution of the upstream medium to the opacity and the FRB dispersion measures, for the parameters considered in the discussion above.

\vspace{1cm}
\section{Discussion and conclusions}\label{sec:discussion}
The claimed detection of a radio flash 76.6 minutes after the sGRB 201006A \citep{rowlinson23}, if real, has significant implications for our understanding of GRBs. The FRB-like radio emission following the purported neutron-star merger is encouraging for follow-up of other GRBs to search for coherent radio emission in their immediate aftermath~\citep[e.g.,][and references therein]{anderson21,tian23}. 

\citet{rowlinson23} conclude that the radio flash associated with \grb{} is from a long-lived neutron star engine born in the aftermath of a neutron-star merger, with the coherent radio emission a product of its eventual demise at $\approx \unit[77]{min}$ (laboratory frame) into a black hole through the Blitzar mechanism~\citep{falcke14}, or an indefinitely stable neutron star producing coherent emission through some mechanism akin to a magnetar flare. 

In this \emph{Letter}, we critically assess both scenarios. In particular, we concretely rule out a neutron-star collapse due to inconsistency with both the observational data of \grb{} and theoretical works for neutron-star collapse times. We also highlight that interpretation of the X-ray afterglow of \grb{} as an indefinitely stable neutron star is subjective to binning. Furthermore, optical constraints require an engine that has significantly accelerated the ejecta, which in turn produces a bright radio afterglow that should currently be observable with the VLA or MeerKAT. While there are caveats to these multi-wavelength constraints, the tuning in the form of rebinning the X-ray afterglow, the optical and radio afterglow non-detections, and the general issues for propagation of coherent radio emission through merger ejecta~\citep{margalit2019_frbbns, Bhardwaj2023} or through magnetospheres~\citep{Beloborodov2022} suggests tension with the interpretation that \grb{} and the potentially associated radio flash are produced by an indefinitely stable neutron star. The tentative, observed dispersion measure of sGRB 201006A is also in significant tension with the expected dispersion measure contributions from the kilonova ejecta at $\unit[49]{min}$, and requires significantly smaller ejected mass than a typical merger~\citep{Radice2024}.

Given the tensions with a neutron-star remnant interpretation, we instead suggest a simpler alternative. That the sGRB was powered by a black hole, with the coherent radio emission a product of a synchrotron maser from a plasmoid interacting with the magnetized upstream material following the sGRB, a scenario that could potentially explain all features of the X-ray data (without needing to rebin) and the optical constraints. Such a scenario also makes predictions for a radio remnant that are not detectable. If this interpretation is correct, then it would serve as great verification for the synchrotron-maser or magnetic reconnection within jet model for coherent radio emission and, in turn, provide evidence that FRB-like phenomena can be associated with objects other than magnetars. 

Alternatively, if the engine is indeed an indefinitely stable neutron star and a kilonova afterglow is not detectable in the near future with radio telescopes such as MeerKAT, then it likely requires that the bulk of the rotational energy is lost via gravitational-wave emission from an r-mode or bar-mode instability. Such strong gravitational-wave emission from an oscillation mode has strong implications on the microphysics of such newly-born neutron stars~\citep{AnderssonReview}, with several implications on the FRB emission mechanism, relativistic jet launching, the expected counterparts to gravitational-wave events, and on gravitational-wave sources for next-generation observatories.
We strongly encourage future radio follow-up to search for the kilonova afterglow of this event to ascertain the nature of the engine, as well as continued monitoring campaigns of other sGRBs to detect coherent radio emission, alongside deep all-sky surveys such as VLASS to detect the kilonova afterglows of other GRBs.


\section{Acknowledgments}
We are grateful to Antonia Rowlinson, Iris De Ruiter, Kelly Gourdji, and Gemma Anderson for helpful discussions and the referee for their comments on the manuscript. N.S. is supported by a Nordita Fellowship. Nordita is supported in part by NordForsk. N.S acknowledges support from the Knut and Alice Wallenberg foundation through the “Gravity Meets Light” project. T. A. C. and S. J. M. receive support from the Australian Government Research Training Program.
P.D.L is supported through Australian Research Council (ARC) Centre of Excellence CE170100004, Discovery Projects DP220101610 and DP230103088, and LIEF Project LE210100002. B.D.M. is supported by the National Science Foundation (grant AST-2002577) and by NASA through the Fermi Guest Investigator Program (grant 80NSSC22K1574).
This work made use of data supplied by the UK Swift Science Data Centre at the University of Leicester.

\bibliographystyle{aasjournal} 
\bibliography{paper}
\end{document}